\documentclass[12pt]{article}
\usepackage[margin=2.6cm]{geometry}
\usepackage{amsmath}
\usepackage[utf8]{inputenc}
\usepackage{scrextend}
\usepackage{url}
\usepackage{bm}
\usepackage{pdflscape}
\usepackage{booktabs}
\usepackage{amsfonts}
\usepackage{graphicx}
\usepackage{xcolor}
\usepackage{array}
\newcolumntype{H}{>{\setbox0=\hbox\bgroup}c<{\egroup}@{}}
\usepackage{listings}
\usepackage[hidelinks]{hyperref}
\providecommand{\keywords}[1]{\textbf{\textit{Keywords:}} #1}

\def\bSig\mathbf{\Sigma}

\newtheorem{thm}{Theorem}
\newtheorem{lem}{Lemma}

\usepackage{amssymb,amsmath}
\usepackage[figuresright]{rotating}
\usepackage{enumerate}
\usepackage{subfigure}
\usepackage{verbatim}
\usepackage{natbib}
\usepackage{multirow}
\usepackage{graphicx}
\usepackage{color}
\usepackage{url}

\newcommand{\TP}{\mathrm{TP}}
\newcommand{\FP}{\mathrm{FP}}
\newcommand{\AUC}{AUC}
\newcommand{\ROC}{ROC}
\newcommand{\AUCm}{\mathrm{AUC}}
\newcommand{\ROCm}{\mathrm{ROC}}
\newcommand{\R}{\texttt{R}}
\newcommand{\mypkg}{\texttt{intcensroc}}
\graphicspath{ {.} }

\title{Predictive Accuracy of Markers or Risk
Scores for Interval Censored Survival Data}

\author{Yuan Wu, Xiaofei Wang, Jiaxing Lin, Beilin Jia,
and Kouros Owzar\\
\small Department of Biostatistics and Bioinformatics, Duke University Medical Center\\
\small Durham, North Carolina, U.S.A.
}
\date{}

\begin{document}

\maketitle

\begin{abstract}
  Methods for the evaluation of the predictive accuracy of biomarkers with
  respect to survival outcomes subject to right censoring have been discussed
  extensively in the literature. In cancer and other diseases, survival outcomes
  are commonly subject to interval censoring by design or due to the follow up
  schema.  In this paper, we present an estimator for the area under the
  time-dependent receiver operating characteristic (\ROC{}) curve for interval
  censored data based on a nonparametric sieve maximum likelihood approach. We
  establish the asymptotic properties of the proposed estimator,
  and illustrate its finite-sample properties using a simulation study. The
  application of our method is illustrated using data from a cancer clinical
  study. An open-source R package to implement the proposed method is available
  on CRAN.
\end{abstract}


\section{ Introduction}

The receiver operator curve (\ROC{})~\cite[]{Zweig:93} serves as an established
and widely used tool for visual assessment of the sensitivity and specificity of
diagnostic tests, and of risk and prediction scores derived from machine
learning applications~\cite[]{spackman:89}. While originally developed for
binary outcomes, the \ROC{} concept has been extended to the evaluation of the
predictive accuracy of tests and markers with respect to time-to-event
outcomes~\citep[]{heagerty2000}. In this context, a time-dependent analog of the
\ROC{} is constructed on the basis of corresponding time-dependent analogs of
sensitivity and specificity.

\par

There exists a rich body of literature for estimation of the time-dependent
\ROC{} when the time-to-event outcome is subject to a right-censoring
mechanism~\cite[]{heagerty2000, heagerty2005, paramita2013}. The application of
these methods is limited in many cancer studies as the corresponding
time-to-event outcomes of interest are invariably subject to interval-censoring
mechanisms by virtue of clinical practice or study design. To be more specific,
we consider CALGB 30801 which is a randomized double-blinded phase III study
evaluating the role of selective COX-2 inhibition in patients with advanced
non-small cell lung cancer~\cite[]{calgb:30801}. The progression of the tumors
in these patients is monitored on the basis of radiologic assessments once every
two months in the first two years and then once every six months in the next
three years. Per protocol, these continue until the first confirmed assessment
of tumor progression. The actual time of this event is not observable as it is
realized between two consecutive assessments dates.  Similarly toxicity events
for this and most cancer studies are not reported in real time but rather by
dosing cycles. Neutropenia, defined as an abnormally low count of neutrophils,
is a common toxicity associated with many chemotherapy agents, including
gemcitabine, pemetrexed or carboplatin, the agents used in CALGB 30801. A
high-grade neutropenia event is defined when the absolute neutrophil count falls
below 1000 cells per microliter of blood. As neutrophil counts are typically
measured right before the chemotherapy dose is administered, the time of this
event, when the neutrophil count crosses the critical threshold, is not
observable. What are observed are the date of the first cycle at which the
patient's count as observed to be below the threshold, along with the date of
the previous cycle when the count was recorded to be above the
threshold. Consequently the actual time of toxicity is not observable and
effectively interval censored between the dates of two consecutive drug cycles.

\par

A na{\"i}ve approach for estimating the \ROC{} in presence of an interval-censored
mechanism is to impute the event time using for example the midpoint or the
right end of the last observed time interval. While this approach is convenient,
in the sense that it allows for re-purposing methods developed for right-censored
data, it is biased.  \cite{li:ma:2011} proposed a non-parametric approach for
estimating the \ROC{} and \AUC{} in presence of interval censoring. To estimate the
curve at time say $t>0$, they exclude the data from any patients whose last
observed time interval contains $t$. The authors formally quantify the loss of
information using a fraction and point out that as this fraction increases, the
accuracy of the estimator decreases and the variance is inflated.  \cite{jac:16}
proposed two approaches to estimate the time-dependent \ROC{} and \AUC{} in the
context of semi-competing time-to-event outcomes subject to interval
censoring. Their first approach is fully model based method, which is based on
the well known Cox regression type illness-death model for the mark effect on
the two competing events. Their second approach also needs the result of illness-death
model result to impute the probability of subjects become diseased before
time $t$ when $t$ is interval censored. These two approaches could be
potentially applied for interval censored single event case. But obviously the
mis-specification of the illness-death model is very likely to introduce
estimation bias in the single event case.

\par

In this paper, we propose a non-parametric approach for estimating the
time-dependent \ROC{} and \AUC{} when the outcome is subject to an
interval-censoring mechanism. Our approach is summarized as follows.  Let $T$
denote the time of the event of interest and $M$ denote a quantitative marker
whose predictive performance with respect to $T$ is to be assessed on the basis
of the time-dependent \ROC{}. $T$ is subject to interval censoring and $M$ is
assumed to be observable.

\par

We first adopt a sieve spline approach to estimate the joint distribution of
$(T,M)$ and the corresponding marginal distribution of $M$. That is, the joint
and marginal functions are restricted in spline function classes (sub-sets of
nonparametric function classes) for their maximum likelihood estimation (MLE).
The resulting estimates are then used to produce plug-in estimates of the
time-dependent sensitivity and specificity functions which are in turn used to
produce a plug-in estimator of the time-dependent \ROC{} function.

\par

The paper is organized as follows. In the next section, we outline the technical
considerations for our proposed method. Thereafter, we illustrate its
finite-sample operating characteristics using a simulation study. Finally, we
present an analysis assessing the performance of COX2 and pgem1 as predictive
biomarkers for progression-free survival (PFS) in advanced non-small cell lung
cancer based on data from CALGB 30801 and conclude the paper with a
discussion. The theoretical results are developed in Web Appendix
1. Specifically, we show that for each $t$ in the support of the censoring time,
this plug-in \ROC{} estimator is uniformly consistent on the support of the
continuous marker, and for each $t$ the corresponding \AUC{} estimator is
consistent. An open-source \R ~\cite[]{r:cite} extension package,
\mypkg{}~\cite[]{intcensroc}, to implement the proposed method is available
through the Comprehensive \R{} Archive Network (CRAN). The scripts to replicate
the results from the simulation study using this package are included as online
supplementary material.

 \section{Methods}
\subsection{Sieve Estimators for the \ROC{} curve and the corresponding \AUC{} }

In this section, we outline a spline-based sieve MLE approach for estimation of
the joint distribution of the event time $T$ and marker $M$. Once this estimate
is obtained, we construct plug-in estimators for the \ROC{} curve and the \AUC{}
at time $t>0$ based on the following definitions as given in
\cite{heagerty2000}:

\begin{equation}
  \label{eq:roc:auc}
  \ROCm{}_t(p)={\TP{}}_t\left\{\FP{}_t^{-1}(p)\right\} \text{~and~}
  \AUCm{}_t=\int_0^1\ROCm{}_t(p)dp,
\end{equation}

where

\begin{equation*}
  \TP{}_t(m)=\frac{F(t,\tau_m)-F(t,m)}{F(t,\tau_m)} \text{~and~}
  \FP{}_t(m)=\frac{1-F_2(m)-F(t,\tau_m)+F(t,m)}{1-F(t,\tau_m)},
\end{equation*}

and where $F(\cdot,\cdot)$ and $F_2(\cdot)$ denote the joint distribution
function of $(T,M)$ and the marginal distribution function of $M$ respectively.

\par

It is supposed that the event time $T$ is interval censored by observation times
$U$ and $V$ and that the marker $M$ is observable.  $(U,V)$ is assumed to be
independent of $(T,M)$.  What is observed for patient $i\in\{1,\ldots,n\}$ is
the sextuple $(u_i,v_i,m_i,\delta^{(1)}_i,\delta^{(2)}_i,\delta^{(3)}_i)$ where
$u_i$ and $v_i$ are the observation times, $m_i$ is the observed marker value,
and $\delta^{(1)}_i=1_{[t_i\leq u_i]}$, $\delta^{(2)}_i=1_{[u_i< t_i\leq v_i]}$
and $\delta^{(3)}_i=1_{[t_i>v_i]}$ are the event indicators for left, interval
and right censoring respectively.  In these event definitions, $t_i$ denotes the
latent event time for patient $i$. Note that $U$ and $V$ could be two random
observation times or result from a group censoring mechanism.  The reader is
referred to \cite{sun:06} for a detailed account on interval censoring.

\par

By virtue of the independence assumption between $(U, V)$ and $(T, M)$, the
likelihood is reduced to

\begin{align}
  \label{like}
\Pi_{i=1}^n\left\{\frac{\partial F(u_i,m_i)}{\partial
m}\right\}^{\delta^{(1)}_i}\left\{\frac{\partial F(v_i,m_i)}{\partial
m}-\frac{\partial F(u_i,m_i)}{\partial m}\right\}^{\delta^{(2)}_i}\left\{\frac{d
F_2(m_i)}{d m}-\frac{\partial F(v_i,m_i)}{\partial m}\right\}^{\delta^{(3)}_i}.
\end{align}

As discussed in~\cite{wu:zhang:12}, a purely nonparametric MLE approach for
optimizing (\ref{like}) is both computationally and theoretically challenging.
To optimize the likelihood, we propose to use spline-based sieve approach.
Suppose $T\in[0,\tau_t]$ and $M\in[0,\tau_m]$ where $\tau_t$ and $\tau_m$ are
two fixed constants.  Construct two sets of B-splines of order $l$
\cite[]{schumaker:81}: $\{B_j^{(1),l}(t)\}_{j=1}^{p_n}$ with knot sequence
$\tilde{\xi}$ as
\begin{align*}
\tilde{\xi}=\{&(\xi_j)_{j=1}^{p_n+l}:\\
&0=\xi_1=\cdots=\xi_l<\xi_{l+1}<\cdots<\xi_{p_n}<\xi_{p_n+1}=\xi_{p_n+l}=\tau_t\},
\end{align*}
and $\{B_k^{(2),l}(m)\}_{k=1}^{q_n}$  with the knot sequence $\tilde{\eta}$ as
\begin{align*}
\tilde{\eta}=\{&(\eta_k)_{k=1}^{q_n+l}:\\
&0=\eta_1=\cdots=\eta_l<\eta_{l+1}<\cdots<\eta_{q_n}<\eta_{q_n+1}=\eta_{q_n+l}=\tau_m\},
\end{align*}
where $p_n$ and $q_n$ are both positive integers dependent on the
sample size $n$.  Let

\begin{equation}\label{Fspline}
F_n(t,m)=\sum_{j=1}^{p_n}\sum_{k=1}^{q_n}\alpha_{j,k}B^{(1),l}_j(t)B^{(2),l}_k(m)
\end{equation}

and

\begin{equation}\label{F2spline}
F_{n,2}(m)=\sum_{k=1}^{q_n}\beta_{k}B^{(2),l}_k(m),
\end{equation}

be the joint and marginal distribution functions for $(T,M)$ restricted to
classes of spline functions. As discussed in~\cite{wu:zhang:12}, by the fact
that $F_n(0,0)=F_{n,2}(0)=0$ as distribution functions, the constraints for
spline coefficients are given as

\begin{equation}\label{consb1}
\begin{split}
&\alpha_{j,1}=0 \text{ for } j=1,\ldots,p_n,\\
&\alpha_{1,k}=0 \text{ for } k=2,\ldots,q_n,\\
&(\alpha_{j+1,k+1}-\alpha_{j+1,k})-(\alpha_{j,k+1}-\alpha_{j,k})\geq 0 \\
&\hspace{15mm}\text{ for } j=1,\ldots,p_n-1, k=1,\ldots,q_n-1,\\
&\beta_1=0, \\
&(\beta_{k+1}-\beta_k)-(\alpha_{p_n,k+1}-\alpha_{p_n,k})\geq0 \text{ for } k=1,\ldots,q_n-1,\\
&\beta_{q_n}\leq1.
\end{split}
\end{equation}

Substitute $F_n(\cdot,\cdot)=F(\cdot,\cdot)$ and $F_{n,2}(\cdot)=F_2(\cdot)$ in
(\ref{like}), by (\ref{Fspline}) and (\ref{F2spline}) we obtain the following
spline-based log likelihood function

\begin{equation}\label{likeb}
\begin{split}
  &\bar{l}_n(\mathbf{\alpha},\mathbf{\beta};)=
  \sum_{i=1}^{n}\left[\delta_{i}^{(1)}\log\frac{\partial\sum_{j=1}^{p_n}\sum_{k=1}^{q_n}\alpha_{j,k}B_{j}^{(1),l}(u_i)B_{k}^{(2),l}(m_i)}
    {\partial m}\right.\\
  &\left.+\delta_{i}^{(2)}\log\frac{\partial\left\{\sum_{j=1}^{p_n}\sum_{k=1}^{q_n}\alpha_{j,k}B_{j}^{(1),l}(v_i)B_{k}^{(2),l}(m_i)
        -\sum_{j=1}^{p_n}\sum_{k=1}^{q_n}\alpha_{j,k}B_{j}^{(1),l}(u_i)B_{k}^{(2),l}(m_i)\right\}}{\partial m}\right.\\
  &\left.+\delta_{i}^{(3)}\log\frac{\partial\left\{\sum_{k=1}^{q_n}\beta_{k}B_{k}^{(2),l}(m_i)
        -\sum_{j=1}^{p_n}\sum_{k=1}^{q_n}\alpha_{j,k}B_{j}^{(1),l}(v_i)B_{k}^{(2),l}(m_i)\right\}}{\partial
      m}\right],
\end{split}
\end{equation}

where
$\mathbf{\alpha}=\left\{\alpha_{j,k}\right\}_{j=1,\cdots,p_n,k=1,\cdots,q_n}$
and $\mathbf{\beta}=\left\{\beta_{k}\right\}_{k=1,\cdots,q_n}$.  In the proposed
sieve MLE approach, nonparametric distribution functions are restricted to
classes of spline functions for their estimation. This is equivalent to finding
the maximizer $(\hat{\mathbf{\alpha}}, \hat{\mathbf{\beta}})$ for (\ref{likeb})
subject to the constraints in (\ref{consb1}).
By plugging  $(\hat{\mathbf{\alpha}},\hat{\mathbf{\beta}})$ into
(\ref{Fspline}) and (\ref{F2spline}) we obtain the sieve MLE for
$(F_0(\cdot,\cdot), F_{0,2}(\cdot))$, the true distribution functions for $(T,M)$.

 \subsection{Computing the Sieve MLE}\label{sec:com}

Given that the B-spline based sieve MLE approach for (\ref{likeb}) involves
complicated constraints (\ref{consb1}). Similar to the approach used
by~\cite{wu:zhang:12}, we propose to use I-splines and its derivatives to
simplify the computation.

\par

Let $I_j^l$ and $M_j^l$ be I-spline and M-spline, respectively, as defined by
\cite{ramsay} and \cite{schumaker:81}, where $M_j^l(t)=\frac{dI_j^l(t)}{dt}$.
\cite{wu:zhang:12} showed that $I_j^l(t)=\sum_{h=j+1}^{p_n+1}B_h^{l+1}(t)$. Note
that $I_j^l$ is of degree $l$, both $N_j^l$ and $B_j^l$ are of degree $l-1$.  By
some algebra we see that $F_{n}(\cdot,\cdot)$ and $F_{n,2}(\cdot)$ given by
(\ref{Fspline}) and (\ref{F2spline}) with constraints (\ref{consb1}) are
equivalent to
\begin{equation}\label{FIspline}
F_n(t,m)=\sum_{j=1}^{p_n-1}\sum_{k=1}^{q_n-1}\gamma_{j,k}I^{(1),l}_j(t)I^{(2),l}_k(m),
\end{equation}
and
\begin{equation}\label{F2Ispline}
F_{n,2}(m)=\sum_{k=1}^{q_n-1}\left\{\sum_{j=1}^{p_n-1}\gamma_{j,k}+\omega_{k}\right\}I^{(2),l}_k(m).
\end{equation}

subject to the constraints
\begin{equation}\label{consi}
\begin{split}
&\gamma_{j,k}\geq 0 \text { for } j=1,\cdots, p_n-1 , k=1,\cdots,q_n-1,\\
&\omega_k\geq 0, k=1,\ldots,q_n-1,\\
&\sum_{j=1}^{p_n-1}\sum_{k=1}^{q_n-1}\gamma_{j,k}+\sum_{k=1}^{q_n-1}\omega_k\leq1.
\end{split}
\end{equation}

By (\ref{FIspline}) and (\ref{F2Ispline}) and $M_j^l(t)=\frac{dI_j^l(t)}{dt}$,
we rewrite the B-spline-based log likelihood (\ref{likeb}) as
\begin{equation}\label{likei}
\begin{split}
&\bar{l}_n(\mathbf{\gamma},\mathbf{\omega})=
\sum_{i=1}^{n}\left[\delta_{i}^{(1)}\log\left\{\sum_{j=1}^{p_n-1}\sum_{k=1}^{q_n-1}
\gamma_{j,k}I_{j}^{(1),l-1}(u_i)M_{k}^{(2),l-1}(m_i)\right\}
\right.\\
&\left.+\delta_{i}^{(2)}\log\left\{\sum_{j=1}^{p_n-1}\sum_{k=1}^{q_n-1}\gamma_{j,k}I_{j}^{(1),l-1}(v_i)M_{k}^{(2),l-1}(m_i)
-\sum_{j=1}^{p_n-1}\sum_{k=1}^{q_n-1}\gamma_{j,k}I_{j}^{(1),l-1}(u_i)M_{k}^{(2),l-1}(m_i)\right\}\right.\\
&\left.+\delta_{i}^{(3)}\log\left\{\sum_{k=1}^{q_n-1}\left(\sum_{j=1}^{p_n-1}\gamma_{j,k}+\omega_{k}\right)M^{(2),l}_k(m_i)
-\sum_{j=1}^{p_n-1}\sum_{k=1}^{q_n-1}\gamma_{j,k}I_{j}^{(1),l-1}(v_i)M_{k}^{(2),l-1}(m_i)\right\}\right],
\end{split}
\end{equation}
where $\mathbf{\gamma}=\left\{\gamma_{j,k}\right\}_{j=1,\cdots,p_n-1,k=1,\cdots,q_n-1}$, $\mathbf{\omega}=\left\{\omega_{k}\right\}_{k=1,\cdots,q_n-1}$.

Now, the proposed sieve MLE problem is equivalent to finding the maximizer
$\left(\hat{\mathbf{\gamma}}, \hat{\mathbf{\omega}}\right)$ for (\ref{likei})
subject to the simpler set of constraints (\ref{consi}). The optimization can be
efficiently implemented using the generalized gradient projection
algorithm~\cite[]{jam:04, zhang:hua:huang,wu:zhang:12}.

\par

The spline knot sequence for the event time component is chosen based on the
observed times $\{(u_i,v_i)\}_{i=1}^n$. Specifically, we first let
$\mathcal{O}=\left\{u_i\delta_i^{(1)}+\frac{u_i+v_i}{2}\delta_i^{(2)}+v_i\delta_i^{(3)}\right\}_{i=1}^n$,
that is, each member of $\mathcal{O}$ equals $u_i$ for left censoring,
$(u_i+v_i)/2$ for interval censoring and $v_i$ for right censoring.
Then we let the number of the interior knots be $[n^{1/3}]$ (the closest integer
to $n^{1/3}$), and put interior knots at the quantiles of $\mathcal{O}$. In the
marker direction, the knot sequence can be directly chosen based on the
quantiles of
$\{m_i\}_{i=1}^n$.

\par

As we have pointed in Section 2.1, once we have the sieve MLE estimates, the
plug-in spline estimators for \ROC{} and \AUC{} are readily obtained
(see~(\ref{eq:roc:auc})). For statistical inference for the \AUC{}, we propose
to use the BCa method~\cite[]{Bradley:96} for computing bootstrap confidence
intervals.

 \section{Simulation Study}

We evaluate the finite-sample operating characteristics of our method on the
basis of the following simulation study.  We assume that event time $T$ follows
an exponential distribution with hazard rate $\lambda>0$ and that the marker $M$
follows a beta distribution with density
$f_2(m)=\frac{\Gamma(\alpha+\beta)}{\Gamma(\alpha)\Gamma(\beta)}
m^{\alpha-1}(1-m)^{\beta-1}$.  The joint distribution of $(T,M)$ is assumed to
be generated by a Clayton copula (\cite{nelsen:06}), with parameter $\mu>1$
\begin{equation*}
  F(t,m)=\Pr(T<t,M<m)=\left\{F_1(t)^{\mu-1}+F_2(m)^{\mu-1}-1\right\}^{1/(\mu-1)},
\end{equation*}
where $F_1(\cdot)$ denotes the marginal distribution function of $T$, and as denoted in Section 2, $F_2(\cdot)$ represents the marginal distribution function of $M$.  We quantify the
dependence between $T$ and $M$ using Kendall's $\tau$ \cite[]{daniel:90}.  Note
that for Clayton's copula larger values of the dependence parameter $\mu$ imply
stronger association. More specifically, $\mu$ is related to $\tau$ through
$\tau=\frac{\mu-1}{\mu+1}$ \cite[]{nelsen:06}.

\par

The number of assessments, $K_c$, is assumed to follow a geometric distribution
with parameter $\nu>0$. The distance between two contiguous assessment times,
$L_c$, is assumed to be fixed. For a given right censoring rate $\rho\in (0,1)$,
the parameter $\nu$ is calibrated so that $\rho=\Pr(T>L_cK_c)$.  Uniform noise,
distributed over the interval $(-L_c/6,L_c/6)$, is added to each assessment time
to account for patient non-compliance.  Based on the relationship between event
time and the actual assessment times we can get the actual values for
$u_i,v_i,\delta_i^{(1)},\delta_i^{(2)}$ and $\delta_i^{(3)}$ in the likelihood
function (\ref{likei}). Note that $\delta_i^{(1)}=1$ (left censoring) implies
$u_i=L_c$, $\delta_i^{(2)}=1$ (interval censoring) implies $u_i$ and $v_i$ are
two consecutive observation times with $v_i-u_i=L_c$ and $\delta_i^{(3)}=1$
(right censoring) implies $v_i$ is the last assessment time.

\par

For estimation, we consider spline basis functions of order $l=3$, that is, we
use quadratic and M-spline basis functions, cubic I-spline basis functions
throughout the simulation as mentioned in Section~\ref{sec:com}.  The knot
sequence for the splines is chosen as described in Section~\ref{sec:com}.  $T$
is generated from an exponential distribution with hazard rate
$\lambda=\log(2)/30$. $M$ is generated from a beta distribution with
$\alpha=2.35$ and $\beta=1.87$, and then $M$ is scaled from $0$ to $10$.  We
consider $\tau=0.2$ and $\tau=0.6$ to represent weak versus strong association,
and right censoring rates of $\rho=0.3$ and $\rho=0.5$ to represent low versus
high levels of censoring. We choose $L_c=6$ for assessment times. The \ROC{} and
the \AUC{} are estimated at times $t=12$ and $28$.  We consider sample sizes of
$n=100$ and $300$.  Coverage probabilities, at the nominal two-sided 95\% level, are
assessed by calculating confidence intervals using the BCa method on the basis of
$B=1000$ bootstrap replicates.  Each illustration is based on $N=1000$
simulation replicates. We note that the putative parameter values for the
distributions are chosen to mimic those from CALGB 30801. 

\par

The relative bias (re-Bias), standard deviation and coverage probability, at the
nominal two-sided 95\% level, for estimation of \AUC{} are shown in
Table~\ref{tab:1}. We observe that for the scenarios we have considered, the
relative bias is less than 6\% for $n=100$ and less than 2\% for $n=300$.  Our
approach provides consistent coverage, at the nominal two-sided confidence level
of 95\%, when $n=300$. We note that strong association seemingly results in
larger bias. We also note that the bias is larger at time point $t=12$ under a
right censoring rate of $0.3$ than under a right censoring rate of $0.5$.  We
will comment on these two issues in the discussion.

\begin{table}[htbp]
\centering\caption{The proposed sieve estimation for \AUC{}}\label{tab:1}
\begin{tabular}{ccccccc}
\hline
Time & True \AUC{} & Right censoring rate & Size & re-Bias & Std &95$\%$ CP\\
  \hline
  \multicolumn{7}{c}{$\tau=0.2$}\\
  \hline
  \multirow{4}{*}{12}&\multirow{4}{*}{0.6818}&\multirow{2}{*}{0.3}&100 &-0.0130&0.0629 &0.845 \\
  &&&300 &-0.0031 &0.0365&0.961 \\
   \cline{3-7}
  &&\multirow{2}{*}{0.5}&100 &-0.0011 &0.0628&0.878 \\
  &&&300 &-0.0030 &0.0384& 0.935\\
  \hline
  \multirow{4}{*}{28}&\multirow{4}{*}{0.6397}&\multirow{2}{*}{0.3}&100 &-0.0090&0.0552  &0.846 \\
  &&&300& -0.0006&0.0323& 0.932\\
 \cline{3-7}
  &&\multirow{2}{*}{0.5}&100 &-0.0217 &0.0518&0.840 \\
  &&&300& -0.0009&0.0355 &0.932\\
  \hline
   \multicolumn{7}{c}{$\tau=0.6$}\\
  \hline
   \multirow{4}{*}{12}&\multirow{4}{*}{0.9473}&\multirow{2}{*}{0.3}&100 &-0.0347 &0.0270& 0.960\\
  &&&300 &-0.0161 &0.0137&0.961 \\
   \cline{3-7}
  &&\multirow{2}{*}{0.5}&100 &-0.0368 &0.0243&0.954\\
  &&&300 & -0.0101 &0.0160&0.968\\
  \hline
   \multirow{4}{*}{28}&\multirow{4}{*}{0.8948}&\multirow{2}{*}{0.3}&100 & -0.0173&0.0314 &0.954 \\
  &&&300&-0.0020&0.0194 & 0.930\\
  \cline{3-7}
  &&\multirow{2}{*}{0.5}&100 &-0.0570&0.0380 & 0.816\\
  &&&300& -0.0092&0.0210 &0.953\\
  \hline
\end{tabular}
\end{table}

\begin{figure}[htbp]
\begin{center}
  \scalebox{0.5}{\includegraphics[angle =0]{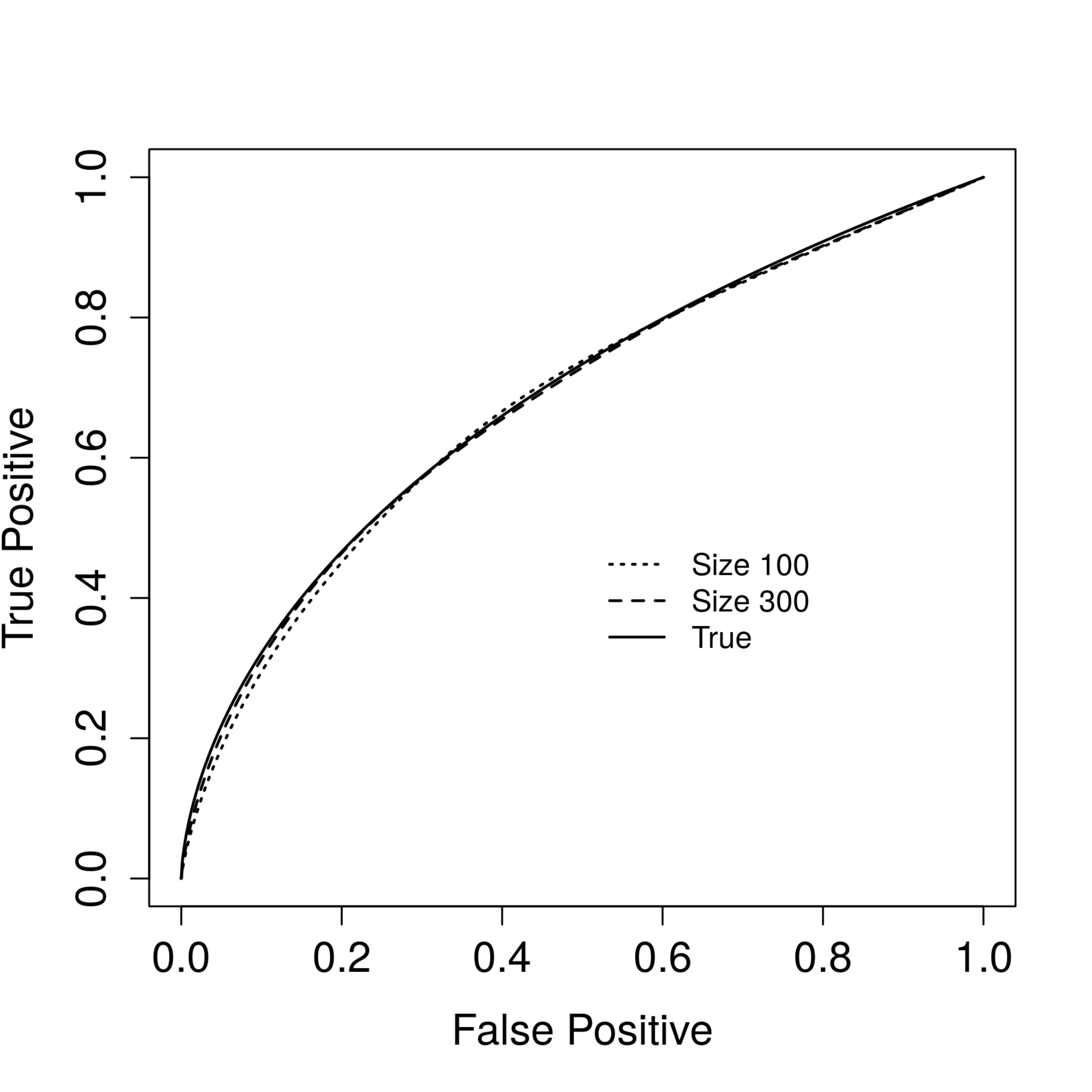}}
\caption{The average sieve estimated \ROC{} curves for two sample sizes, $n=100$ and $300$,
  with a right censoring rate of 0.5 and $\tau=0.2$ evaluated at time $t=12$}
\label{fig:1}
\end{center}
\end{figure}

\begin{figure}[htbp]
\begin{center}
  \scalebox{0.5}{\includegraphics[angle =0]{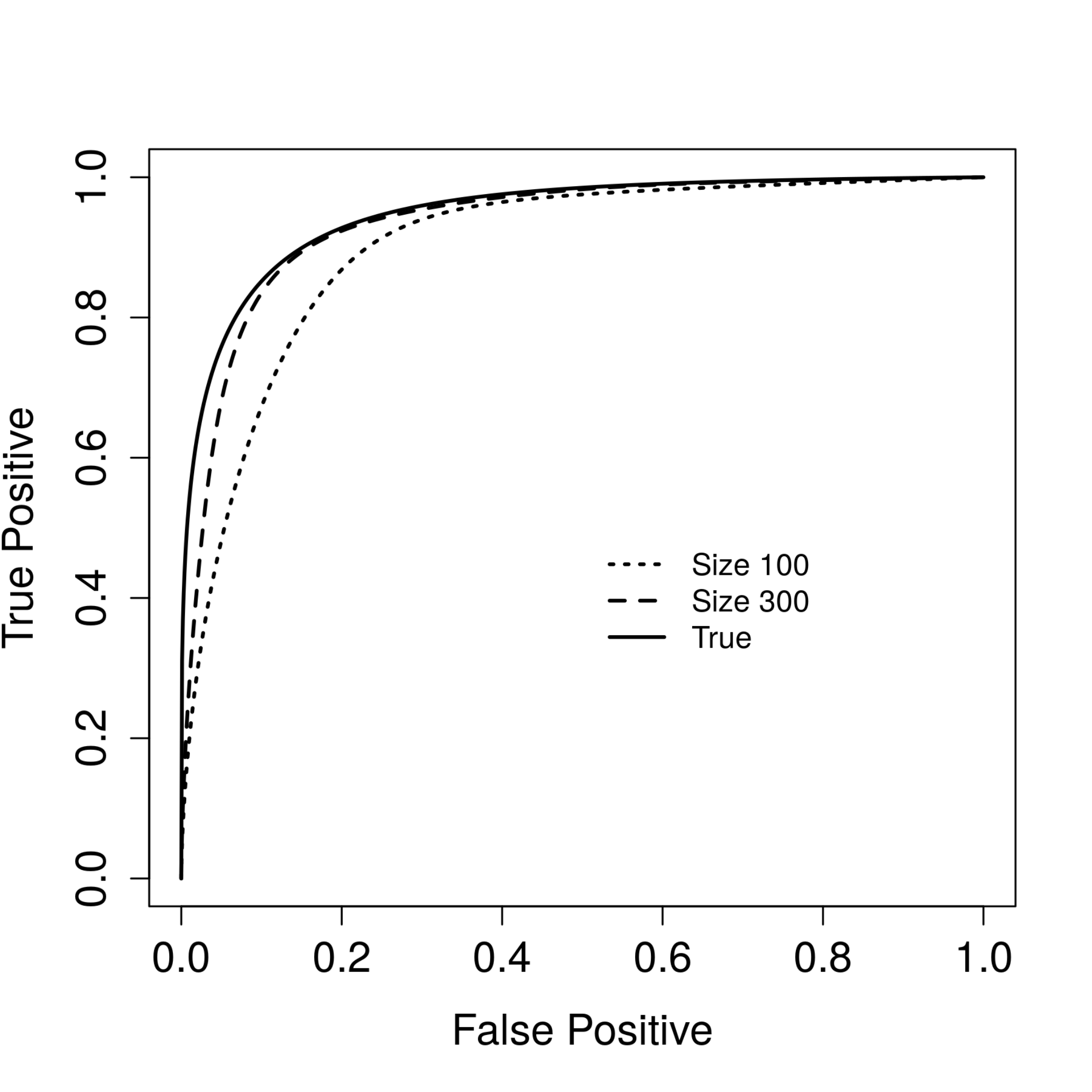}}
  \caption{The average sieve estimated \ROC{} curves for two sample sizes,
    $n=100$ and $300$, with a right censoring rate of 0.5 and $\tau=0.6$ evaluated at
    time $t=12$}
\label{fig:2}
\end{center}
\end{figure}

Figure~\ref{fig:1} and Figure~\ref{fig:2} present the results for estimating the
\ROC{} curve, at time $t=12$ under a right censoring rate of $0.5$ for
$\tau=0.2$ and $0.6$, and $n=100$ and $300$. The estimation becomes more accurate when the
size is increased from $100$ to $300$, as expected, and less accurate as the
association becomes stronger, which is consistent with the results in
Table~\ref{tab:1}.

 \section{Analysis of CALGB 30801}

\begin{figure}[htbp]
\begin{center}
            \scalebox{0.7}{\includegraphics[angle =0]{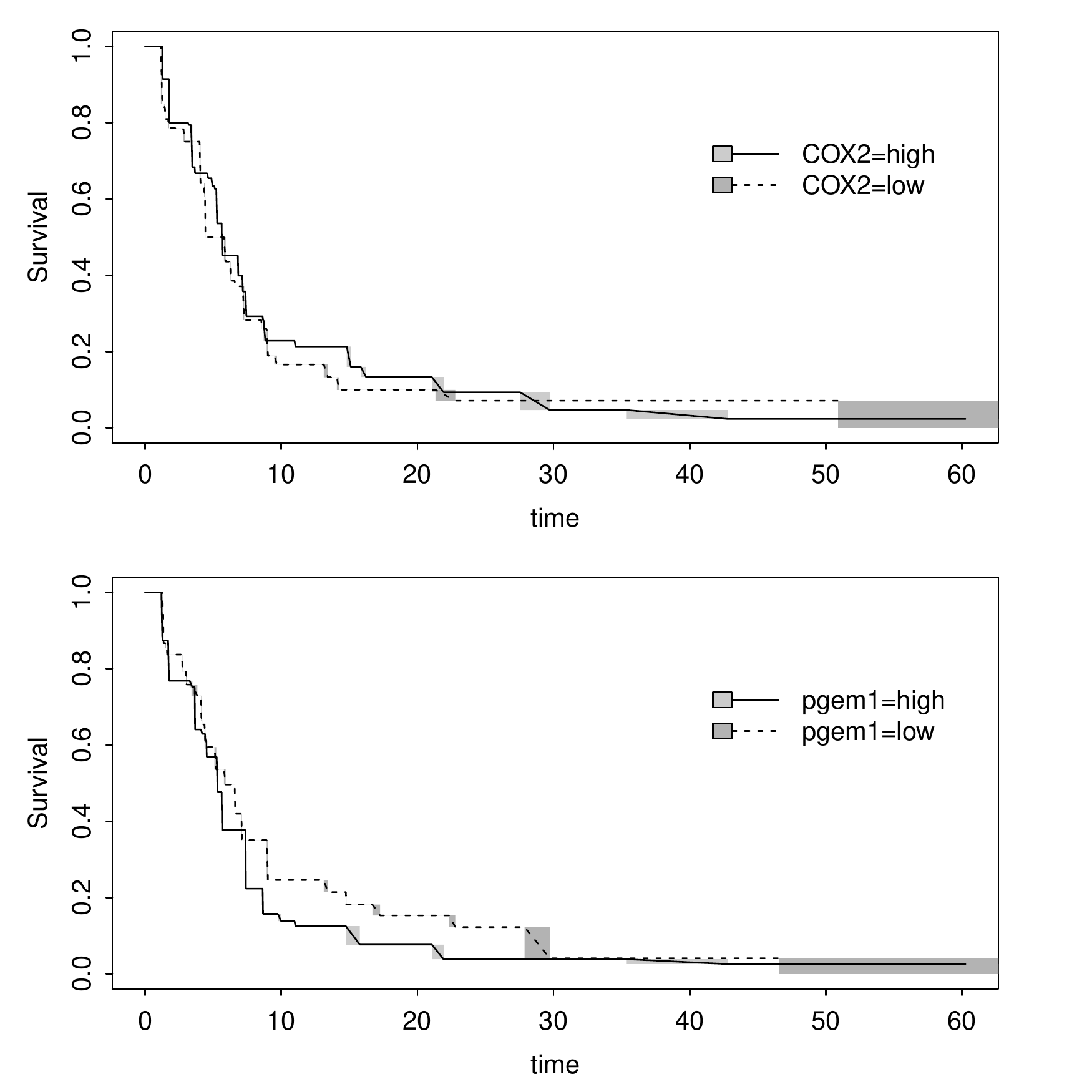}}
\caption{Kaplan-Meier plots for CALGB 30801  data for COX-2 and pgem1
  markers, marker levels are separated into two groups regarding to their median
  levels as ``high" and ``low" category.}\label{fig:KMreal}
\end{center}
\end{figure}

We applied our \AUC{} estimator for analysis of CALGB 30801 data (randomized
phase III double blind trial evaluating selective COX-2 inhibition in COX-2
expressing advanced non-small cell lung cancer). The CALGB 30801 data includes
interval censored progression free survival and two markers (COX-2 and pgem1)
for 312 patients. The median survival time is 10.9 weeks. We also produce
the Kaplan-Meier plots (Figure~\ref{fig:KMreal}) for both COX-2 and pgem1
markers, the markers are transferred into a binary factor ``low" and ``high"
regarding to the median marker levels.

\par

For the purpose of demonstration, we only include patients with observed
markers, and we treated  patients without
progression events after the last follow-up visit  as right
censored in our \AUC{} estimator.

\par

The two \AUC{} estimates for two markers are $0.50$ and $0.55$ with $95\%$
confidence interval $[0.4721, 0.5139]$ and $[0.5000, 0.6498]$ for  time
at 12 weeks, respectively.  The confidence intervals are obtained by bootstrap,
the number of resample are $10000$ for both markers. The bootstrap confidence
intervals are computed using BCa method \cite[]{Bradley:96}. Since both \AUC{}
values are not significantly greater that $0.5$, neither marker is very helpful
to predict the event time.

 \section{Concluding Remarks}

We have proposed spline based plug-in estimators for time dependent \ROC{} curve
and its corresponding \AUC{} measure based on interval censored time to event
data and continuous marker.  Our simulation studies show very good performance,
with respect to bias, for our proposed method with practical finite sample
sizes. The results also suggest the BCa bootstrapping confidence interval
can be used for statistical inference on our proposed \AUC{} estimator
when the sample size is large.

\par

Two observations from the simulation results shown in Table~\ref{tab:1}
bear discussion.

\par
Comparing the results for $\tau=0.2$ and $\tau=0.6$, we observe that stronger
association between event time and marker seemingly increases relative bias.
This is likely due to the fact that the two knot sequences, for estimating the
marginal distributions of $T$ and $M$, were chosen independently. The
suboptimality of this approach is likely to become more pronounced as the
association between $T$ and $M$ becomes stronger. This explanation also applies
to the results for estimating the \ROC{} curve shown in Figure~\ref{fig:1}
and Figure~\ref{fig:2} where the discrepancy between the actual \ROC{} curve and its
estimates is larger for stronger association.

\par

\begin{figure}[htbp]
\begin{center}
  \scalebox{0.5}{\includegraphics[angle =0]{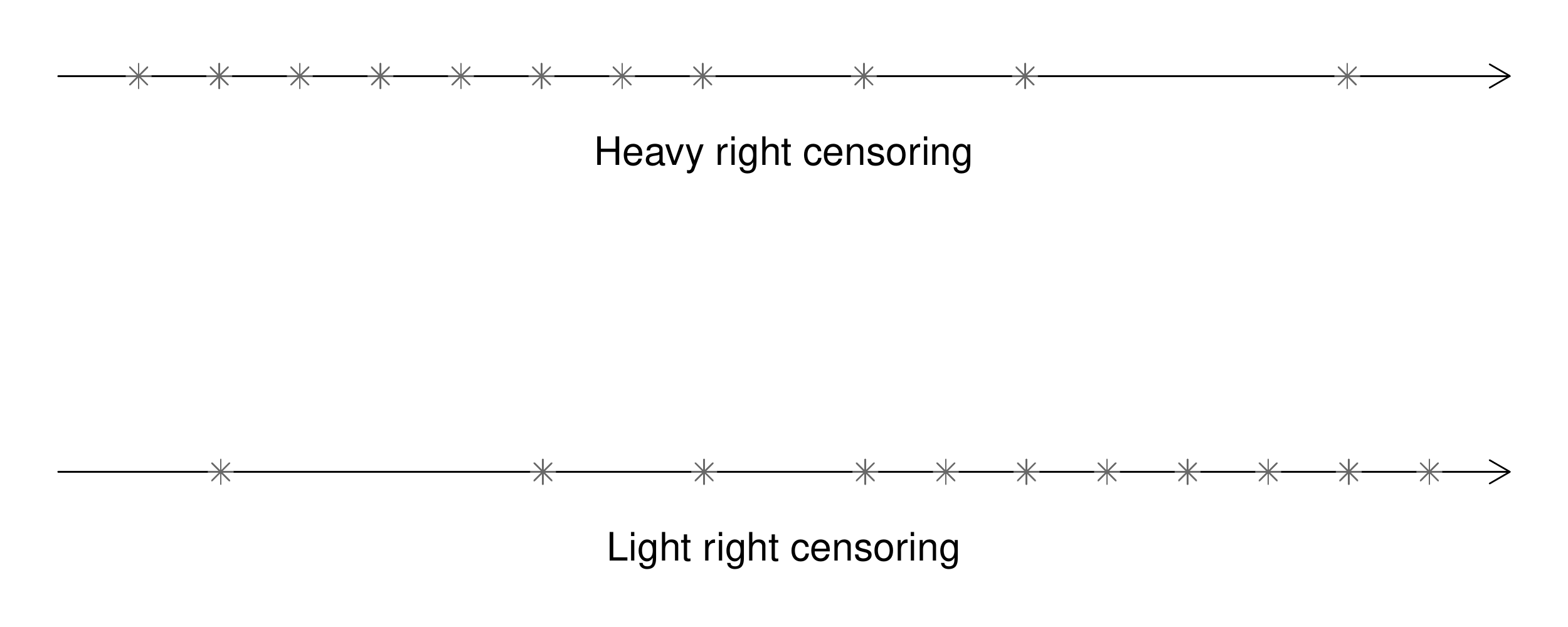}}
  \caption{The distributions of the current status times under heavy and light
    right censoring}
\label{fig:3}
\end{center}
\end{figure}

When estimating \AUC{} at time $12$ (a relatively early time), the relative bias
under a right censoring rate of $0.3$ is larger than that under a right
censoring rate of $0.5$.  In Figure~\ref{fig:3}, we illustrate the distributions
of current status times under light and heavy right censoring. We observe that
under light right censoring, the distribution of the current status times is
skewed to the left away from the early time points. As the knots are assigned
based on quantiles of the censoring times, the performance of sieve estimation
at relatively early times may be worse under light right censoring than that at
later time points.

\par

Besides simulation, we have established the consistency of our estimators theoretically (see Web
Appendix 1). The proposed method can be extended to the competing risk or
multi-marker frameworks.  A user-friendly \R~\cite[]{r:cite} package to
implement our proposed method has already been released on CRAN
(\url{https://CRAN.R-project.org/package=intcensROC}).

\vspace{10pt}
\centerline{\Large{\sc Acknowledgements}}

\noindent
The research was supported in part by award number P01CA142538 from the
National Cancer Institute. The content is solely the responsibility of the
authors and does not necessarily represent the official views of the National
Institute of Health.

\section*{Asymptotic Properties}

We establish asymptotic consistency of the plug-in sieve estimators for the
target \ROC{} curve and \AUC{}. Study of the asymptotic properties needs
empirical process theory and requires some regularity conditions on the joint
and marginal distributions of the marker and event time. Let $F_0(\cdot,\cdot)$
denote true joint distribution function for $(T,M)$, and $F_{0,2}(\cdot)$ denote
the true marginal distribution function for $M$, respectively.  The following
conditions sufficiently guarantee the results in the forthcoming Theorem
\ref{thm:1}.

\subsection*{Regularity Conditions:}
\begin{enumerate}[C1.]

\item $\frac{\partial^2 F_0(t,m)}{\partial t\partial m}$ has a positive lower
  bound in $[0,\tau_t]\times [0,\tau_m]$.

\item All $p$th mixed partial derivatives of
  $\frac{\partial F_0(t,m)}{\partial m}$ are continuous in
  $[0,\tau_t]\times [0,\tau_m]$; The $p$th derivative of
  $\frac{dF_{0,2}(m)}{dm}$ is continuous on $[0,\tau_m]$.

\item The random observation times $U$ and $V$ are both in $[\tau_1, \tau_2]$,
  with $\tau_1>0, \text{ and } \tau_2<\tau_t$ with $V-U\ge \tau_0$ for fixed
  $\tau_0, \tau_1$ and $\tau_2$.

\item $(U,V)$ either has discrete distribution or a continuous probability
  density function (pdf). If $(U,V)$ is discrete, its probability mass function
  has a positive lower bound. Otherwise, the pdf of $(U,V)$ has a positive lower
  bound.

\end{enumerate}

Let $\|\cdot\|_{L_r(P_{U,M})}$, $\|\cdot\|_{L_r(P_{V,M})}$ and
$\|\cdot\|_{L_r(P_{M})}$ denote the $L_r$-norms associated with the probability
measures $P_{U,M}$, $P_{V,M}$ and $P_{M}$ for $(U,M)$, $(V,M)$ and $M$,
respectively.  For $\theta_1=\left\{f_1(\cdot,\cdot),h_1(\cdot)\right\}$ and
$\theta_2=\left\{f_2(\cdot,\cdot),h_2(\cdot)\right\}$, we define

 \begin{equation}\label{distance}
 d(\theta_1,\theta_2)=\left\{\|f_1-f_2\|_{L_2(P_{U,M})}^2+\|f_1-f_2\|_{L_2(P_{V,M})}^2+\|h_1-h_2\|_{L_2(P_{M})}^2\right\}^{1/2}.
 \end{equation}

Let $X=(U,V,M,\Delta_1,\Delta_2, \Delta_3)$. For a single observation
$x=(u,v,m,\delta^{(1)},\delta^{(2)},\delta^{(3)})$ from $X$ and the variable of
nonparametric functions $\theta=\{(G(\cdot,\cdot), G_2(\cdot)\}$, the log
likelihood function is given by
\begin{align*}
l(\theta;x)=\delta^{(1)}\log G(u,m)+\delta^{(2)}\left\{G(v,m)-G(u,m)\right\}+\delta^{(3)}\left\{G_2(m)-G(v,m)\right\},
\end{align*}
where $G(t,m)=\frac{dF(t,m)}{dm}$ and $G_2(m)=\frac{dF_2(m)}{dm}$, with the
joint and marginal distribution functions $F(\cdot,\cdot)$ for $(T,M)$ and
$F_2(\cdot)$ for $M$.

Denote $\mathbb{M}(\theta)=Pl(\theta;x)$ with $P$ being the true joint
probability measure of $X$, and $\mathbb{M}_n(\theta)=\mathbb{P}_nl(\theta;x)$
with $\mathbb{P}_nf=\frac{1}{n}\sum_{i=1}^n f(x_i)$ the empirical process
indexed by $f(X)$.

Let
$\theta_0=\left(G_0, G_{0,2}\right)\equiv\left(\frac{\partial F_0}{\partial m},
  \frac{d F_{0,2}}{d m}\right)$ and
$\theta_n=(G_n,G_{n,2})=\left(\frac{\partial F_n}{\partial m}, \frac{d
    F_{n,2}}{d m}\right)$ for $F_n$ and $F_{n,2}$ being spline functions as
defined by (3) and (4) in the main manuscript.

In what follows, we prove the consistency for the sieve plug-in point estimators
for \ROC{} and \AUC{} in Theorem \ref{thm:1}. To this end, we first prove the
consistency for the proposed sieve MLE for the joint distribution of $(T,M)$ in
the following lemma.

\begin{lem}\label{lem:1}
Suppose that C1--C4 hold and denote $\tilde{l}_n(\theta_n;)$ as
$\bar{l}_n(\mathbf{\alpha},\mathbf{\beta};)$ for
$\bar{l}_n(\mathbf{\alpha},\mathbf{\beta};)$ defined by (6) in the main
manuscript. Then there exists a class $\Theta_n$ with elements $\theta_n$'s,
for the maximizer $\hat{\theta}_n$ of $\tilde{l}_n(\theta_n;)$ over
$\Theta_n$, we have that $$d(\hat{\theta}_n,\theta_0)\rightarrow_P 0.$$
\end{lem}

Denote $\hat{F}_{n}(t,m)=\int_0^m\hat{G}_n(t,u)du$ and
$\hat{F}_{n,2}(m)=\int_0^m\hat{G}_{n,2}(u)du$ with
$\hat{\theta}_n=\left(\hat{G}_{n},\hat{G}_{n,2}\right)$. Then by their
cumulative definitions in \cite{heagerty2000}, we write estimators for
$\TP{}_{0,t}(m)\equiv\frac{F_0(t,\tau_m)-F_0(t,m)}{F_0(t,\tau_m)}$ and $\FP{}_{0,t}(m)\equiv\frac{1-F_{0,2}(m)-F_0(t,\tau_m)+F_0(t,m)}{1-F_0(t,\tau_m)}$  as
$$\widehat{\TP{}}_{n,t}(m)=\frac{\hat{F}_{n}(t,\tau_m)-\hat{F}_{n}(t,m)}{\hat{F}_{n}(t,\tau_m)}$$
and
$$\widehat{\FP{}}_{n,t}(m)=\frac{1-\hat{F}_{n,2}(m)-\hat{F}_{n}(t,\tau_m)+\hat{F}_{n}(t,m)}{1-\hat{F}_{n}(t,\tau_m)},$$
respectively. Then the sieve estimators for
$\ROCm{}_{0,t}(p)\equiv\TP{}_{0,t}\left\{\FP{}_{0,t}^{-1}(p)\right\}$ and
$\AUCm{}_{0,t}\equiv\int_0^1\ROCm{}_{0,t}(p)dp$
are given as
$$\widehat{\ROCm{}}_{n,t}(p)=\widehat{\TP{}}_{n,t}\left\{\widehat{\FP{}}_{n,t}^{-1}(p)\right\}$$
and $$\widehat{\AUCm{}}_{n,t}=\int_0^1\widehat{\ROCm{}}_{n,t}(p)dp,$$
respectively.

\begin{thm}\label{thm:1}
Suppose that C1--C4 hold. Then for any $t$ in the support of $U$ or $V$
$$\sup_{p\in[0,1]}\left|\widehat{\ROCm{}}_{n,t}(p)-\ROCm{}_{0,t}(p)\right|\rightarrow_P0.$$
In addition, $\widehat{\AUCm{}}_{n,t}$ is a consistent estimator for
$\AUCm{}_{0,t}$  for any $t$ in the support of $U$ or $V$.
\end{thm}

\section*{Technical Proofs}
{\em\underline{Proof of Theorem \ref{thm:1}}}

By Lemma \ref{lem:1} and the definition of  $d(\cdot,\cdot)$ in
(\ref{distance}), $\left\|\hat{G}_n-G_0\right\|_{L_2(U,M)}\rightarrow_P 0$,
$\left\|\hat{G}_n-G_0\right\|_{L_2(V,M)}\rightarrow_P 0$ and
$\left\|\hat{G}_{n,2}-G_{0,2}\right\|_{L_2(M)}\rightarrow_P 0$. By the
properties of $\Theta_n$ and regularity conditions C1, C2 and C4, using the
similar arguments as Lemma 0.7 in \cite{wu:zhang:12} we can establish that for
any $t$ on the support of $U$ or $V$
$$\sup_{m\in[0,\tau_m]}\left|\hat{G}_n(t,m)-G_0(t,m)\right|\rightarrow_P0$$
and
$$\sup_{m\in[0,\tau_m]}\left|\hat{G}_{n,2}(m)-G_{0,2}(m)\right|\rightarrow_P0.$$
Then we can show that $\widehat{\TP{}}_{n,t}(m)\rightarrow_P \TP{}_{0,t}(m)$ and
$\widehat{\FP{}}_{n,t}(m)\rightarrow_P \FP{}_{0,t}(m)$ both uniformly for
$m\in[0,\tau_m]$. It is clear that $\FP{}_{0,t}(m)$ has continuous inverse
function $\FP{}^{-1}_{0,t}(p)$ for any $p\in[0,1]$. Now we show that for
$\widehat{\FP{}}^{-1}_{n,t}(p)$ (the inverse function of
$\widehat{\FP{}}_{n,t}(m)$)
$$\sup_{p\in[0,1]}\left|\widehat{\FP{}}^{-1}_{n,t}(p)-\FP{}^{-1}_{0,t}(p)\right|\rightarrow_P 0 .$$
Since $\FP{}^{-1}_{0,t}(p)$ is continuous at any $p\in[0,1]$, then it is
uniformly continuous on $[0,1]$. If we denote $m_p=\FP{}^{-1}_{0,t}(p)$, then
for any $\epsilon>0$ there exists $\delta>0$ such that $|p'-p|<2\delta$
implies $|\FP{}^{-1}_{0,t}(p')-m_p|<\epsilon$ for any $p\in[0,1]$ and $p'\in[0,1]$ . Hence,
$$\FP{}_{0,t}(m_p-\epsilon)-p>\delta$$
and $$\FP{}_{0,t}(m_p+\epsilon)-p<-\delta.$$ Next suppose
$$\sup_{m\in[0,\tau_m]}\left|\widehat{\FP{}}_{n,t}(m)-\FP{}_{0,t}(m)\right|<\delta.$$
If we also denote $m_n=\widehat{\FP{}}^{-1}_{n,t}(p)$, then we have
$m_n>m_p-\epsilon$. Since if not, then
$$\FP{}_{0,t}(m_p-\epsilon)-p<\widehat{\FP{}}_{n,t}(m_p-\epsilon)+\delta-p\le
\delta,$$ which contradicts the previous inequality. Similarly, we can use
contradiction to show that $m_n<m_p+\epsilon$. Hence, for any $p\in[0,1]$
$$|m_n-m_p|<\epsilon.$$

The preceding arguments imply that
\begin{align*}
\Pr\left(\sup_{p\in[0,1]}\left|\widehat{\FP{}}^{-1}_{n,t}(p)-\FP{}^{-1}_{0,t}(p)\right|\ge\epsilon\right)
&\le\Pr\left(\sup_{m\in[0,\tau_m]}\left|\widehat{\FP{}}_{n,t}(m)-\FP{}_{0,t}(m)\right|\ge\delta\right)\\
&\rightarrow 0,
\end{align*}
by the uniform convergence for $\widehat{\FP{}}_{n,t}(m)$, as we discussed at the beginning of the proof. Hence, we complete the verification for
$$\sup_{p\in[0,1]}\left|\widehat{\FP{}}^{-1}_{n,t}(p)-\FP{}^{-1}_{0,t}(p)\right|\rightarrow_P 0 .$$
Next, it is easily seen that $\TP{}_{0,t}(m)$ is uniformly continuous function in variable $m$ on
$[0,\tau_m]$. Therefore, by the uniform convergence for
$\widehat{\TP{}}_{n,t}(m)$ (as briefly discussed at the beginning of this proof),
and applying the continuous mapping theorem (with continuous mapping $\TP{}_{0,t}(m)$) on the
preceding established
uniform convergence for $\widehat{\FP{}}^{-1}_{n,t}(p)$,
we have
\begin{align*}
\sup_{p\in[0,1]}\left|\widehat{\ROCm{}}_{n,t}(p)-\ROCm{}_{0,t}(p)\right|
=&\sup_{p\in[0,1]}\left|\widehat{\TP{}}_{n,t}\left\{\widehat{\FP{}}^{-1}_{n,t}(p)\right\}-\TP{}_{0,t}\left\{\FP{}^{-1}_{0,t}(p)\right\}\right|\\
\le&\sup_{p\in[0,1]}\left|\widehat{\TP{}}_{n,t}\left\{\widehat{\FP{}}^{-1}_{n,t}(p)\right\}
-\TP{}_{0,t}\left\{\widehat{\FP{}}^{-1}_{n,t}(p)\right\}\right|\\
&+\sup_{p\in[0,1]}\left|\TP{}_{0,t}\left\{\widehat{\FP{}}^{-1}_{n,t}(p)\right\}-\TP{}_{0,t}\left\{\FP{}^{-1}_{0,t}(p)\right\}\right|\\
\rightarrow&_P0.
\end{align*}

The consistency for $\widehat{\AUCm{}}_{n,t}$ is then trivial, that is
$$ \left|\widehat{\AUCm{}}_{n,t}-\AUCm{}_{0,t}\right|
\le\int\limits_0^1\sup_{p\in[0,1]}\left|\widehat{\ROCm{}}_{n,t}(p)-\ROCm{}_{0,t}(p)\right|\to_P 0. ~~\square$$

{\em\underline{Proof of Lemma \ref{lem:1}}}

We apply Theorem 5.7 in \cite{vander:98} to show the consistency.  Following the
proof of this theorem, we need to find a set containing both $\theta_0$ and
$\hat{\theta}_n$ (as set ``$\Theta$" in Theorem 5.7 in \cite{vander:98}).

To find the sub class $\Theta_n$ as addressed in Lemma \ref{lem:1}, we enforce
the following conditions on $(G_n,G_{n,2})=\left(\frac{\partial F_n}{\partial
m}, \frac{d F_{n,2}}{d m}\right)$:

\begin{enumerate}[T1.]
\item
$F_n$ and $F_{n,2}$ satisfy the conditions for a joint distribution function and
a corresponding marginal distribution function in
$[\tau_1,\tau_2]\times[0,\tau_m]$ and on $[0,\tau_m]$, respectively.

\item
$G_n$ and $G_{n,2}$ are defined in $[\tau_1,\tau_2]\times[0,\tau_m]$ and on $[0,\tau_m]$, respectively.
\item
$G_n$ and $\left|\frac{\partial G_n}{\partial t}\right|$, $\left|\frac{\partial
		G_n}{\partial m}\right|$, $\left|\frac{\partial^2 G_n}{\partial
		t^2}\right|$, $\left|\frac{\partial^2 G_n}{\partial m^2}\right|$ and
		$\left|\frac{\partial^2 G_n}{\partial t\partial m}\right|$ all have a
		positive upper bound.
\item
$G_{n,2}$ and $\left|\frac{d G_{n,2}}{d m}\right|$ both have a positive upper bound.
\item
Let $\Delta_{i}^{\xi}=\xi_{i+1}-\xi_i$ and
		$\Delta_{j}^{\eta}=\eta_{j+1}-\eta_j$.  For $i=1,\cdots,p_n-1$ and
		$j=1,\cdots,q_n-1$, $\frac{\min_{i:l\leq i\leq p_n}
		\Delta_i^{\xi}}{\max_{i:l\leq i\leq p_n }\Delta_i^{\xi}}$ and
		$\frac{\min_{j:l\leq j\leq q_n }\Delta_j^{\eta}}{\max_{j:l\leq j\leq q_n}
		\Delta_j^{\eta}}$ both have  positive lower bounds.
\item
$G_n(u,m)$, $G_n(v,m)-G_n(u,m)$ and $G_{n,2}(m)-G_n(v,m)$ all have  positive
		lower bounds for $(u,v)\in[\tau_1,\tau_2]$ with $v-u\ge\tau_0$ and
		$m\in[0,\tau_m]$.
\end{enumerate}

And $\Theta_n$ is defined by
\begin{align}\label{thetan}
\Theta_n=\left\{\theta_n=(G_n,G_{n,2}): \text{T1--T6 hold}\right\}.
\end{align}
Now we create a more general class $\Theta$ compared to $\Theta_n$.  That is,
for functions $G(\cdot,\cdot)$ and $G_2(\cdot)$, we enforce the following
conditions:
\begin{enumerate}[H1.]
\item
$G$ and $G_{2}$ are defined on $[\tau_1,\tau_2]\times[0,\tau_m]$ and $[0,\tau_m]$, respectively.
\item
$G$ and $\left|\frac{\partial G}{\partial t}\right|$, $\left|\frac{\partial
		G}{\partial m}\right|$, $\left|\frac{\partial^2 G}{\partial t^2}\right|$,
		$\left|\frac{\partial^2 G}{\partial m^2}\right|$ and $\left|\frac{\partial^2
		G}{\partial t\partial m}\right|$ all have a positive upper bound.
\item
 $G_{2}$ and $\left|\frac{d G_{2}}{d m}\right|$ both have a positive upper bound.
\item
$G(u,m)$, $G(v,m)-G(u,m)$ and $G_2(m)-G(v,m)$ all have positive lower bounds for
		$(u,v)\in[\tau_1,\tau_2]$ with $v-u\ge\tau_0$ and $m\in[0,\tau_m]$.
\end{enumerate}
Now we define
\begin{align}\label{theta}
\Theta=\left\{\theta=(G,G_{2}):\text{H1--H4 hold}\right\}
\end{align}
It is obvious that  $\Theta_n\subset\Theta$. On the other hand, by regularity
conditions C1 and C2 it can be shown that  $\theta_0\in\Theta$. Hence, both
$\theta_0$ and $\hat{\theta}_n$ are contained in $\Theta$.  In what follows we
complete the proof by verifying the conditions of Theorem 5.7 in
\cite{vander:98}.
First, we verify
$\sup_{\theta\in\Theta}\left|\mathbb{M}_n(\theta)-\mathbb{M}(\theta)\right|\rightarrow_P
0$. Denote $\mathfrak{L}=\{l(\theta;x):\theta\in\Omega\}$.  Since
$$\sup_{\theta\in\Omega}\left|\mathbb{M}_n(\theta)-\mathbb{M}(\theta)\right|
=\sup_{l(\theta;X)\in\mathfrak{L}}\left|(\mathbb{P}_n-P)l(\theta;X)\right|\rightarrow_P
0,$$ it suffices to show that $\mathfrak{L}$ is a $P$-Glivenko-Cantelli.

Let $\Theta_{G}=\left\{G:\theta=(G,G_2), \theta\in\Theta\right\}$ and
$\Theta_{G_2}=\left\{G_2:\theta=(G,G_2), \theta\in\Theta\right\}$. By H2 and the
bracket numbers for Sobolev spaces, we know that there exists
$\parallel\cdot\parallel_\infty \epsilon$-brackets $$\left[G^{L,1},
G^{R,1}\right], \cdots, \left[G^{L,[e^{c/\epsilon}]},
G^{R,[e^{c/\epsilon}]}\right]$$ to cover $\Theta_{G}$. Similarly, by H3 and the
bracket numbers for Sobolev spaces, we know that there exists
$\parallel\cdot\parallel_\infty \epsilon$-brackets $$\left[G_2^{L,1},
G_2^{R,1}\right], \cdots, \left[G_2^{L,[e^{c/\epsilon}]},
G_2^{R,[e^{c/\epsilon}]}\right]$$ to cover $\Theta_{G_2}$.

Hence, it is easy to construct a set of brackets $\left[l_{i,j}^L,
l_{i,j}^R\right]$ with $i=1, \cdots, [e^{c/\epsilon}]$ and
$j=1,\cdots,[e^{c/\epsilon}]$ that for any $l(\theta;x)\in \mathfrak{L}$ with
any observation $x=(t,u,v,q,z,\delta_1,\delta_2,\delta_3)$  we have
$l_{i,j}^L\le l(\theta;x)\le l_{i,j}^R$, where
\begin{align*}
l_{i,j}^L=\delta_1\log G^{L,i}(u,m)+\delta_2\left\{G^{L,i}(v,m)-G^{R,i}(u,m)\right\}+\delta_3\left\{G_2^{L,j}(m)-G^{R,i}(v,m)\right\}
\end{align*}
and
\begin{align*}
l_{i,j}^R=\delta_1\log G^{R,i}(u,m)+\delta_2\left\{G^{R,i}(v,m)-G^{L,i}(u,m)\right\}+\delta_3\left\{G_2^{R,j}(m)-G^{L,i}(v,m)\right\}.
\end{align*}
It can be seen that $\parallel l_{i,j}^R-l_{i,j}^L\parallel_\infty\le c\epsilon$
by some algebra using property H4 for $\Theta$.  This leads to the conclusion
that $N_{[~]}(\epsilon,\mathfrak{L},\parallel\cdot\parallel_\infty)\le
e^{c/\epsilon}$.

Then by $N_{[~]}(\epsilon,\mathfrak{L},L_1(P))\le
N_{[~]}(\epsilon,\mathfrak{L},\parallel\cdot\parallel_\infty)$, we have
$N_{[~]}(\epsilon,\mathfrak{L},L_1(P))\le e^{c/\epsilon}$. Hence, $\mathfrak{L}$
is a $P$-Glivenko-Cantelli by Theorem 2.4.1 in \cite{van:wellner:96}.

Second, by lemma \ref{lem:2}, we have that for any $\theta\in\Theta$,
$$\mathbb{M}(\theta_0)-\mathbb{M}(\theta)\ge cd(\theta,\theta_0)^2.$$

Finally, we verify $\mathbb{M}_n\left(\hat{\theta}_n\right) \ge \mathbb{M}_n(\theta_0) - o_P(1)$.

By regularity conditions C1 and C2, and the construction of $\Theta_n$,
Jackson's Theorem on page 149 in \cite{deboor:01} and Lemma 0.2 in the
supplemental material of \cite{wu:zhang:12} imply that there exists
$\theta_n=(G_n, G_{n,2})$ in $\Theta_n$ such that
$\|G_{n}-G_{0}\|_\infty\leq c(n^{-p\kappa})$ and
$\|G_{n,2}-G_{0,2}\|_\infty\leq c(n^{-p\kappa})$. Since $\hat{\theta}_n$
maximizes $\mathbb{M}_n(\theta)$ in $\Theta_n$,
$\mathbb{M}_n(\hat{\theta}_n)-\mathbb{M}_n(\theta_n)>0.$ Hence,
\begin{align*}
\mathbb{M}_n\left(\hat{\theta}_n\right)-\mathbb{M}_n(\theta_0)
&=\mathbb{M}_n\left(\hat{\theta}_n\right)-\mathbb{M}_n\left(\theta_n\right)+\mathbb{M}_n\left(\theta_n\right)-\mathbb{M}_n(\theta_0)\\
&\ge \mathbb{M}_n\left(\theta_n\right)-\mathbb{M}_n(\theta_0)\\
&=\left(\mathbb{P}_n-P\right)\{l(\theta_n;X)-l(\theta_0;X)\}+P\{l(\theta_n;X)-l(\theta_0;X)\}
\end{align*}
By regularity conditions C1, C2 and C3, and the construction of $\Theta_n$,
using some algebra, we get
$$P\{l(\theta_n;X)-l(\theta_0;X)\}^2\rightarrow
0~~\text{as}~~n\rightarrow\infty.$$ Then
\begin{align*}
\rho_P\{l(\theta_n;X)-l(\theta_0;X)\}&=\left(P\left[\{l(\theta_n;X)-l(\theta_0;X)\}
-P\{l(\theta_n;X)-l(\theta_0;X)\}\right]^2\right)^{1/2}\\
&\le \left[P\{l(\theta_n;X)-l(\theta_0;X)\}^2\right]^{1/2}\rightarrow0~~\text{as}~~n\rightarrow\infty.
\end{align*}

By $N_{[~]}(\epsilon,\mathfrak{L},L_2(P))\le
N_{[~]}(\epsilon,\mathfrak{L},\parallel\cdot\parallel_\infty)$, we have
$N_{[~]}(\epsilon,\mathfrak{L},L_2(P))\le e^{c/\epsilon}$. Then
\begin{align*}
J_{[~]}\left(\delta,\mathfrak{L},L_2(P)\right)&=\int_0^\delta \sqrt{\log
	N_{[~]}\left(\epsilon,\mathfrak{L},L_2(P)\right)}
	d\epsilon\le\int_0^\delta\sqrt{\log\left( e^{c/\epsilon}\right)}d\epsilon\\
&\le\int_0^\delta\sqrt{\left(\frac{c}{\epsilon}\right)}d\epsilon\le
	c\int_0^\delta \epsilon^{-1/2}d\epsilon=c\delta^{1/2} <\infty.\\
\end{align*}
So $\mathfrak{L}$ is Donsker by Theorem 19.5 in \cite{vander:98}. Then by
Corollary 2.3.12 in \cite{van:wellner:96} we have
$$\left(\mathbb{P}_n-P\right)\{l(\theta_n;X)-l(\theta_0;X)\}=o_P\left(n^{-1/2}\right).$$

Furthermore, by the Cauchy-Schwarz inequality
\begin{align*}
|P\{l(\theta_n;X)-l(\theta_0;X)\}|\le P|l(\theta_n;X)-l(\theta_0;X)|\le
	c\left[P\{l(\theta_n;X)-l(\theta_0;X)\}^2\right]^{1/2}\rightarrow0,
\end{align*}
as  $n\rightarrow\infty$.

Then $Pl(\theta_n;X) > l(\theta_0;X) - o(1)$. Hence,
$$\mathbb{M}_n\left(\hat{\theta}_n\right)-\mathbb{M}_n(\theta_0)\ge
o_P\left(n^{-1/2}\right)-o(1)=-o_P(1).$$ This completes the proof of
$d\left(\hat{\theta}_n,\theta\right)\rightarrow_P0$. $~~\square$

\begin{lem}\label{lem:2}
  Given that C1--C4 hold. For any $\theta\in\Theta$ for $\Theta$ defined by
  (\ref{theta}).
$$\mathbb{M}(\theta_0)-\mathbb{M}(\theta)\ge cd(\theta,\theta_0)^2.$$
\end{lem}

{\em\underline{Proof of Lemma \ref{lem:2}}}

For $\theta\in\Theta$, the likelihood function with one observation $x$ is
denoted as
$$L(\theta;x)=G(u,m)^{\delta_1}\left\{G(v,m)-G(u,m)\right\}^{\delta_2}\left\{G_2(m)-G(v,m)\right\}^{\delta_3}.$$
For the vector of true distribution functions $\theta_0$, the likelihood
function $L(\theta_0;x)$ is given similarly.

Let $dP/d\mu=\varrho$ for Lebesgue measure (dominating measure) $\mu$. It is
easy to see $\varrho$ is closely related to $L(\theta_0;X)$ since $P$ is the
joint probability measure of $X$. Then by regularity condistions C1, C2, C3
and C4, and the properties of $\Theta$ and the proof of Lemma 5.35 in
\cite{vander:98}
\begin{align*}
\mathbb{M}(\theta_0)-\mathbb{M}(\theta)&=P\log L(\theta_0;X)-P\log
	L(\theta;X)=P\log\frac{L(\theta_0;X)}{L(\theta;X)}\\
&\ge c\int \left(\sqrt{L(\theta_0;x)}- \sqrt{L(\theta;x)}\right)^2 d\mu\ge c\int
	\left(L(\theta_0;x)- L(\theta;x)\right)^2 \varrho d\mu\\
&= cP\left(L(\theta_0;X)- L(\theta;X)\right)^2.
\end{align*}
Since
\begin{align*}
P\left(L(\theta_0;X)- L(\theta;X)\right)^2&
=P\left[\Delta_1\left\{G_0(U,M)-G(U,M)\right\}^2\right]\\
&+P\left(\Delta_2\left[\left\{G_0(V,M)-G_0(U,M)\right\}-\left\{G(V,M)-G(U,M)\right\}\right]^2\right)\\
&+P\left(\Delta_2\left[\left\{G_{0,2}(M)-G_0(V,M)\right\}-\left\{G_{2}(M)-G(V,M)\right\}\right]^2\right),
\end{align*}
where
\begin{align*}
P\left[\Delta_1\left\{G_0(U,M)-G(U,M)\right\}^2\right]
&=E\left[\Delta_1\left\{G_0(U,M)-G(U,M)\right\}^2\right]\\
&=E\left(E\left[\Delta_1\left\{G_0(U,M)-G(U,M)\right\}^2|U,M\right]\right)\\
&=P_{U,M}\left(\left[F_{0,1}(U)\left\{G_0(U,M)-G(U,M)\right\}^2\right]\right)\\
&\ge cP_{U,M}\left\{G_0(U,M)-G(U,M)\right\}^2,
\end{align*}
\begin{align*}
P&\left(\Delta_2\left[\left\{G_0(V,M)-G_0(U,M)\right\}-\left\{G(V,M)-G(U,M)\right\}\right]^2\right)\\
&=E\left(\Delta_2\left[\left\{G_0(V,M)-G_0(U,M)\right\}-\left\{G(V,M)-G(U,M)\right\}\right]^2\right)\\
&=E\left\{E\left(\Delta_2\left[\left\{G_0(V,M)-G_0(U,M)\right\}-\left\{G(V,M)-G(U,M)\right\}\right]^2|U,V,M\right)\right\}\\
&=P_{U,V,M}\left(\left[\{F_{0,1}(V)-F_{0,1}(U)\}\left[\left\{G_0(V,M)-G_0(U,M)\right\}\
-\left\{G(V,M)-G(U,M)\right\}\right]^2\right]\right)\\
&\ge cP_{U,V,M}\left[\left\{G_0(V,M)-G_0(U,M)\right\}-\left\{G(V,M)-G(U,M)\right\}\right]^2,
\end{align*}
and
\begin{align*}
P&\left(\Delta_3\left[\left\{G_{0,2}(M)-G_0(V,M)\right\}-\left\{G_{2}(M)-G(V,M)\right\}\right]^2\right)\\
&=E\left(\Delta_3\left[\left\{G_{0,2}(M)-G_0(V,M)\right\}-\left\{G_{2}(M)-G(V,M)\right\}\right]^2\right)\\
&=E\left\{E\left(\Delta_3\left[\left\{G_{0,2}(M)-G_0(V,M)\right\}-\left\{G_{2}(M)-G(V,M)\right\}\right]^2|V,M\right)\right\}\\
&=P_{V,M}\left(\left[\{1-F_{0,1}(V)\}\left[\left\{G_{0,2}(M)-G_0(V,M)\right\}-\left\{G_{2}(M)-G(V,M)\right\}\right]^2\right]\right)\\
&\ge cP_{V,M}\left[\left\{G_{0,2}(M)-G_0(V,M)\right\}-\left\{G_{2}(M)-G(V,M)\right\}\right]^2.
\end{align*}
Now we have
\begin{align*}
\mathbb{M}(\theta_0)-\mathbb{M}(\theta)\ge &cP\left\{G_0(U,M)-G(U,M)\right\}^2\\
&+cP\left[\left\{G_0(V,M)-G(V,M)\right\}-\left\{G_0(U,M)-G(U,M)\right\}\right]^2\\
&+cP\left[\left\{G_{0,2}(M)-G_{2}(M)\right\}-\left\{G_0(V,M)-G(V,M)\right\}\right]^2.
\end{align*}
Since $a^2+b^2\ge (a+b)^2/2$ and $ a^2+b^2+c^2\ge (a+b+c)^2/3$, we have
\begin{align*}
\mathbb{M}(\theta_0)-\mathbb{M}(\theta)\ge &cP\left\{G_0(U,M)-G(U,M)\right\}^2\\
&+cP\left[\left\{G_0(V,M)-G(V,M)\right\}-\left\{G_0(U,M)-G(U,M)\right\}\right]^2\\
\ge&cP\left\{G_0(V,M)-G(V,M)\right\}^2,
\end{align*}
and
\begin{align*}
\mathbb{M}(\theta_0)-\mathbb{M}(\theta)\ge &cP\left\{G_0(U,M)-G(U,M)\right\}^2\\
&+cP\left[\left\{G_0(V,M)-G(V,M)\right\}-\left\{G_0(U,M)-G(U,M)\right\}\right]^2\\
&+cP\left[\left\{G_{0,2}(M)-G_{2}(M)\right\}-\left\{G_0(V,M)-G(V,M)\right\}\right]^2\\
\ge&cP\left\{G_{0,2}(M)-G_{2}(M)\right\}^2.
\end{align*}
We conclude that
\begin{align*}
\mathbb{M}(\theta_0)-\mathbb{M}(\theta)\ge &cP\left\{G_0(U,M)-G(U,M)\right\}^2
+cP\left\{G_0(V,M)-G(V,M)\right\}^2\\
&+cP\left\{G_{0,2}(M)-G_{2}(M)\right\}^2\\
\ge& cd^2(\theta_0,\theta). ~~\square
\end{align*}


\begin{thebibliography}{}

\bibitem[Daniel(1990)Daniel]{daniel:90}
Daniel, W.~W. (1990).
\newblock {\em Applied nonparametric statistics. 2nd ed.}
\newblock PWS-KENT, Boston.

\bibitem[de~Boor(2001)de~Boor]{deboor:01}
de~Boor, C. (2001).
\newblock {\em A Practical Guide to Splines, Revised Ed.}
\newblock Springer, New York.

\bibitem[Edelman {\em et~al.}(2017)Edelman, Wang, Hodgson, Cheney, Baggstrom,
  and Thomas]{calgb:30801}
Edelman, M.~J., Wang, X., Hodgson, L., Cheney, R.~T., Baggstrom, M.~Q., and
  Thomas, S.~P. (2017).
\newblock Phase iii randomized, placebo-controlled, double-blind trial of
  celecoxib in addition to standard chemotherapy for advanced non–small-cell
  lung cancer with cyclooxygenase-2 overexpression: Calgb 30801 (alliance).
\newblock {\em Journal of Clinical Oncology\/}, {\bf 35}, 2184--2192.

\bibitem[Heagerty and Zheng(2005)Heagerty and Zheng]{heagerty2005}
Heagerty, P.~J. and Zheng, Y. (2005).
\newblock Survival model predictive accuracy and roc curves.
\newblock {\em Biometrics\/}, {\bf 61}, 92--105.

\bibitem[Heagerty {\em et~al.}(2000)Heagerty, Lumley, and Pepe]{heagerty2000}
Heagerty, P.~J., Lumley, T., and Pepe, M.~S. (2000).
\newblock Time dependent roc curves for censored survival data and a diagnostic
  marker.
\newblock {\em Biometrics\/}, {\bf 56}, 337--344.

\bibitem[Jacqmin-Gadda {\em et~al.}(2016)Jacqmin-Gadda, Blanche, Chary,
  Touraine, and Dartigues]{jac:16}
Jacqmin-Gadda, H., Blanche, P., Chary, E., Touraine, C., and Dartigues, J.
  (2016).
\newblock Receiver operating characteristic curve estimation for time to event
  with semicompeting risks and interval censoring.
\newblock {\em Statistical Methods in Medical Research\/}, {\bf 25},
  2750--2766.

\bibitem[Jamshidian(2004)Jamshidian]{jam:04}
Jamshidian, M. (2004).
\newblock On algorithms for restricted maximum likelihood estimation.
\newblock {\em Computational Statistics and Data Analysis\/}, {\bf 45},
  137--157.

\bibitem[Li and Ma(2011)Li and Ma]{li:ma:2011}
Li, J. and Ma, S. (2011).
\newblock Time-dependent roc analysis under diverse censoring patterns.
\newblock {\em Statistics in Medicine\/}, {\bf 30}, 1266--1277.

\bibitem[Lin {\em et~al.}(2018)Lin, Wu, Wang, and Owzar]{intcensroc}
Lin, J., Wu, Y., Wang, X., and Owzar, K. (2018).
\newblock {\em intcensROC: Fast Spline Function Based Constrained Maximum
  Likelihood Estimator for AUC Estimation of Interval Censored Survival
  Data\/}.
\newblock R package version 0.1.1.

\bibitem[Nelsen(2006)Nelsen]{nelsen:06}
Nelsen, R.~B. (2006).
\newblock {\em An Introduction to Copulas. 2nd edition\/}.
\newblock Springer, New York.

\bibitem[{R Core Team}(2018){R Core Team}]{r:cite}
{R Core Team} (2018).
\newblock {\em R: A Language and Environment for Statistical Computing\/}.
\newblock R Foundation for Statistical Computing, Vienna, Austria.

\bibitem[Ramsay(1988)Ramsay]{ramsay}
Ramsay, J.~O. (1988).
\newblock Monotone regression splines in action.
\newblock {\em Statist Science\/}, {\bf 3}, 425--441.

\bibitem[Saha-Chaudhuri and Heagerty(2013)Saha-Chaudhuri and
  Heagerty]{paramita2013}
Saha-Chaudhuri, P. and Heagerty, P.~J. (2013).
\newblock Non-parametric estimation of a time-dependent predictive accuracy
  curve.
\newblock {\em Biostatistics\/}, {\bf 14}, 42--59.

\bibitem[Schumaker(1981)Schumaker]{schumaker:81}
Schumaker, L. (1981).
\newblock {\em Spline Function: Basic Theory\/}.
\newblock John Wiley, New York.

\bibitem[Spackman(1989)Spackman]{spackman:89}
Spackman, K.~A. (1989).
\newblock Signal detection theory: Valuable tools for evaluating inductive
  learning.
\newblock In {\em Proceedings of the Sixth International Workshop on Machine
  Learning\/}, pages 160--163, San Mateo, CA.

\bibitem[Sun(2006)Sun]{sun:06}
Sun, J. (2006).
\newblock {\em The Statistical Analysis of Interval-censored Failure Time
  Data\/}.
\newblock Springer-Verlag, New York.

\bibitem[Thomas and Bradley(1996)Thomas and Bradley]{Bradley:96}
Thomas, D.~J. and Bradley, E. (1996).
\newblock Bootstrap confidence intervals.
\newblock {\em Statistical Science\/}, {\bf 11}, 189--212.

\bibitem[van~der Vaart(1998)van~der Vaart]{vander:98}
van~der Vaart, A.~W. (1998).
\newblock {\em Asymptotic Statistics\/}.
\newblock Cambridge University Press, Cambridge.

\bibitem[van~der Vaart and Wellner(1996)van~der Vaart and
  Wellner]{van:wellner:96}
van~der Vaart, A.~W. and Wellner, J.~A. (1996).
\newblock {\em Weak Convergence and Empirical Processes\/}.
\newblock Springer, New York.

\bibitem[Wu and Zhang(2012)Wu and Zhang]{wu:zhang:12}
Wu, Y. and Zhang, Y. (2012).
\newblock Partially monotone tensor spline estimation of the joint distribution
  function with bivariate current status data.
\newblock {\em Annals of Statistics\/}, {\bf 40}, 1609--1636.

\bibitem[Zhang {\em et~al.}(2010)Zhang, Hua, and Huang]{zhang:hua:huang}
Zhang, Y., Hua, L., and Huang, J. (2010).
\newblock A spline-based semiparametric maximum likelihood estimation for the
  cox model with interval-censored data.
\newblock {\em Scandinavian Journal of Statistics\/}, {\bf 37}, 338--354.

\bibitem[Zweig and Campbell(1993)Zweig and Campbell]{Zweig:93}
Zweig, M.~H. and Campbell, G. (1993).
\newblock Receiver-operating characteristic (roc) plots: a fundamental
  evaluation tool in clinical medicine.
\newblock {\em Clinical Chemistry\/}, {\bf 39}, 561--577.

\end{thebibliography}
\end{document}